\documentclass[showpacs,prl,aps,superscriptaddress,preprintnumbers]{revtex4}

\usepackage{epsfig}
\usepackage{graphicx}
\usepackage{amsmath}
\usepackage{amsfonts}
\usepackage{epstopdf}

\def\slashchar#1{\setbox0=\hbox{$#1$}     		
   \dimen0=\wd0                                 	
   \setbox1=\hbox{/} \dimen1=\wd1               	
   \ifdim\dimen0>\dimen1                        	
      \rlap{\hbox to \dimen0{\hfil/\hfil}}      	
      #1                                        	
   \else                                        	
      \rlap{\hbox to \dimen1{\hfil$#1$\hfil}}   	
      /                                         	
   \fi}

\renewcommand{\vec}{\boldsymbol}
\newcommand{\be}{\begin{equation}}
\newcommand{\ee}{\end{equation}}
\newcommand{\bear}{\begin{eqnarray}}
\newcommand{\eear}{\end{eqnarray}}
\newcommand{\ba}{\begin{array}}
\newcommand{\ea}{\end{array}}

\begin{document}

\title{Chiral Electronics}

\author{Dmitri E. Kharzeev}
\email{Dmitri.Kharzeev@stonybrook.edu}
\affiliation{Department of Physics and Astronomy, Stony Brook University, Stony Brook, New York 11794-3800, USA}
\affiliation{Department of Physics,
Brookhaven National Laboratory, Upton, New York 11973-5000, USA}

\author{Ho-Ung Yee}
\email{hyee@uic.edu}
\affiliation{Department of Physics, University of Illinois, 
Chicago, Illinois 60607, USA}
\affiliation{
RIKEN-BNL Research Center, Brookhaven National Laboratory, 
Upton, New York 11973-5000, USA}


\begin{abstract}
We consider the properties of electric circuits involving Weyl semimetals. The existence of the anomaly-induced chiral magnetic current 
in a Weyl semimetal subjected to magnetic field causes an interesting and unusual behavior of such circuits. 
We consider two explicit examples: i) a circuit involving the ``chiral battery" and ii) a circuit that can be used as a ``quantum amplifier" of magnetic field. 
The unique properties of these circuits stem from the chiral anomaly and may be utilized for creating ``chiral electronic" devices.
\end{abstract}

\maketitle
Recently, the 3D materials with linearly dispersing excitations \cite{Wan:2011} have attracted significant attention.  The existence of these ``chiral" excitations stems from the point touchings of conduction and valence bands. The corresponding dynamics is described by the Hamiltonian 
$
H = \pm v_F \vec{\sigma} \cdot \vec{k},
$ 
where $v_F$ is the Fermi velocity of the quasi-particle, $\vec{k}$ is the momentum in the first Brillouin zone, and $\vec{\sigma}$ are the Pauli matrices. 
This Hamiltonian describes massless particles with positive or negative (depending on the sign) chiralities, e.g. neutrinos, and the corresponding wave equation is known as the Weyl equation -- hence the name {\it Weyl semimetal} \cite{Wan:2011}. Weyl semimetals are closely related to 2D graphene \cite{Geim:2007}, and to the topological insulators \cite{TI} -- 3D materials with a gapped bulk and a surface supporting chiral excitations.  Specific realizations of Weyl semimetals have been proposed, including a multilayer structure composed of identical thin films of a magnetically doped 3D topological insulator, separated by ordinary-insulator spacer layers \cite{wsm}. 

Weyl semimetals provide a unique opportunity to study the macroscopic behavior of systems composed by chiral fermions. In particular, they allow \cite{cme} to study, in a condensed matter system, the {\it chiral magnetic  effect} expected 
\cite{Kharzeev:2004ey, Kharzeev:2007tn, Kharzeev:2007jp, Fukushima:2008xe, Kharzeev:2009fn}, and possibly observed experimentally at Relativistic Heavy Ion Collider \cite{Abelev:2009ac}, in chirally imbalanced quark-gluon plasma in the presence of an external magnetic effect as a consequence of axial anomaly in QCD$\times$QED. Closely related phenomena have been discussed in the physics of neutrinos  \cite{Vilenkin:1979ui}, conductors with mirror isomer symmetry \cite{elia,leonid}, primordial electroweak plasma \cite{Giovannini:1997gp} and quantum wires \cite{acf}.  Note that the role of axial anomaly and the corresponding Chern-Simons dynamics are crucial for the existence of the chiral magnetic current; without the anomaly, this current has to vanish in thermal equilibrium, in contrast to naive arguments. The effects of the anomaly on the transport in Weyl semimetals, including the chiral magnetic effect, have recently been investigated in \cite{sy,Zahed:2012yu,Son:2012bg,zyuzin}.

In this letter, we would like to consider some of the electric circuits involving Weyl semimetals. We will argue that the existence of chiral magnetic current 
in Weyl semimetal subjected to magnetic field can cause an interesting, and potentially useful for practical applications, behavior of such circuits. 
To be specific, we will consider two explicit examples: i) a circuit involving the {\it chiral battery} \cite{Fukushima:2008xe}; and ii) a circuit that can be used as an amplifier of magnetic field, possibly opening a way to creating a sensor of ultra-weak magnetic fields. 

 \begin{figure}[t]
 \includegraphics[scale=0.2]{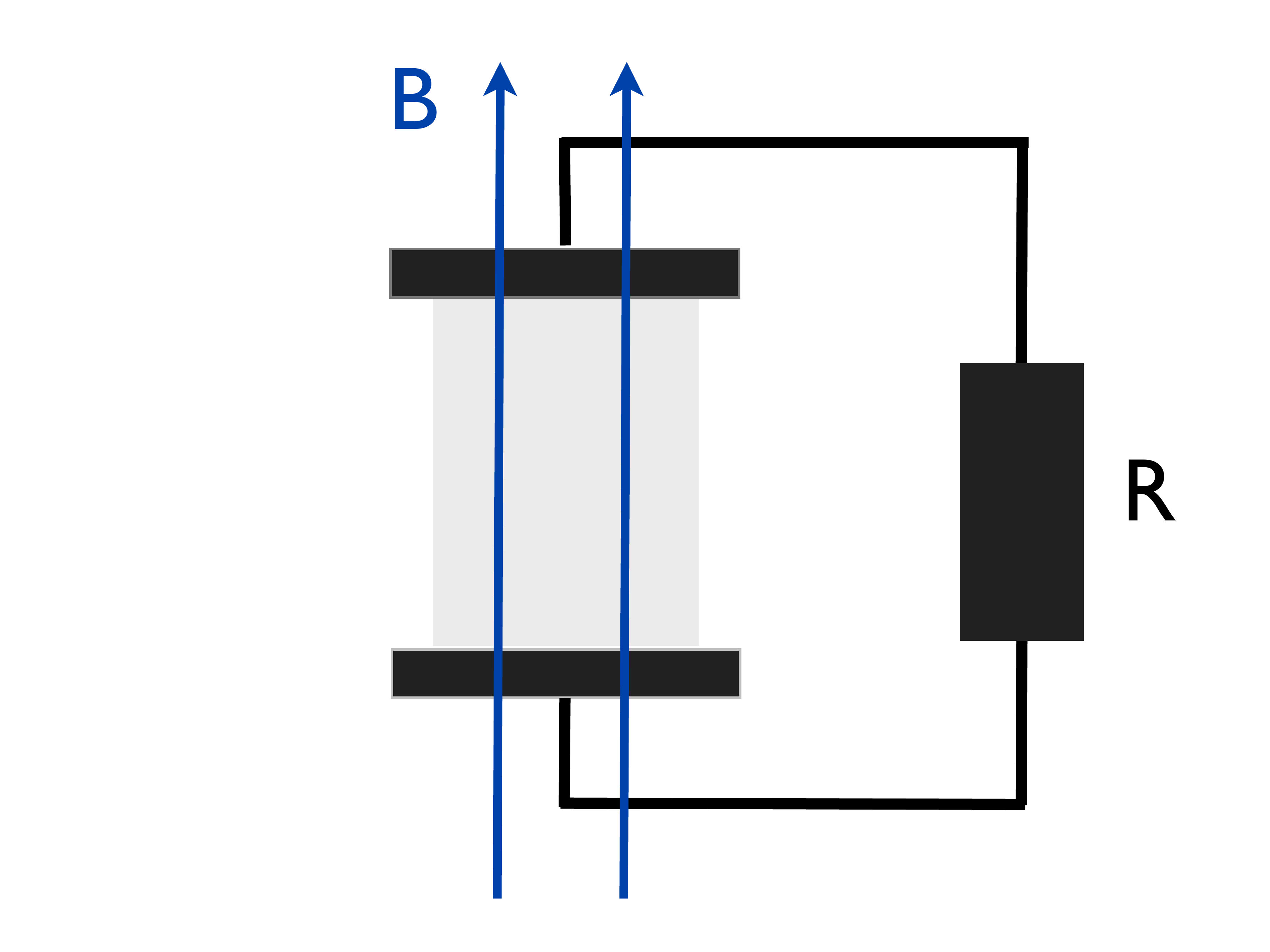}
 \caption{
 The chiral battery: Weyl semimetal (shown in grey) connected to the circuit with resistance $R$ in an external magnetic field $B$.
 }
 \label{sol1}
 \end{figure}

Consider first a cylindrical sample of Weyl semimetal put inside a solenoid that provides an external, constant magnetic field of strength $B$ along the
longitudinal direction of cylindrical geometry, say $\hat x^3$, see Fig. 1.
The top and bottom of the Weyl semimetal are touching metallic plates that can conduct electric 
currents flowing through the sample. These two metallic plates are then connected to an outside circuit 
which is characterized by a resistance $R$. Let the cross section area of the Weyl semimetal sample be $A$ and
the longitudinal length be $d$.

When an external magnetic field is present, there exists an anomaly-induced chiral magnetic current density along $\hat x^3$, given by 
\be
J_{a}={e^2\over 4\pi^2 } \sum_i k_i \mu_i  B \quad,\label{ia}
\ee
where $\mu_i$ is the chiral chemical potential specifying the difference between the chemical potential of excitations with opposite chiralities, and 
the sum is over different kinds of band touching Weyl points (``flavors" of chiral fermions). The chiral chemical potential can either be a property of the Weyl 
semimetal  \cite{cme}, or can be stored in the system through the anomaly equation in parallel electric and magnetic fields \cite{Fukushima:2008xe}. However in the former case it is a conserved quantity (for a fixed geometry of the sample) and thus does not give rise to the chiral magnetic current in equilibrium. 
Once an external magnetic field is applied, the energy stored in the difference of the chemical potentials of left- and right-handed fermions can be released 
by generating the current (\ref{ia}), hence the name {\it chiral battery} \cite{Fukushima:2008xe}.

Note that according to (\ref{ia}) a Weyl semimetal is a kind of battery that provides a definite amount of current,
contrary to conventional batteries that support a definite voltage. 
The total anomaly-induced current through the sample is $I_a=A J_a$, where $A$ is the area.  
However, this is not the entire current. Let the entire current be $I$, then there is a voltage drop along the resistance $R$
given by $\Delta V=IR$. Since the same amount of voltage drop should also occur along the Weyl semimetal sample,
there is an electric field along $\hat x^3$ direction with a magnitude
\be
E=-{\Delta V\over d}=-{IR\over d}\quad;\label{electric}
\ee
note the negative sign of $E$. This electric field gives rise to a normal current through conductivity $\sigma$
\be
I_n=A\sigma E=-{A\sigma R\over d} I=-{R\over R_0} I\quad,\label{in}
\ee
where $R_0\equiv {d\over \sigma A}$ is the intrinsic resistance of the Weyl semimetal sample.
The total current $I$ should be the sum of $I_a$ (\ref{ia}) and $I_n$ (\ref{in}), that is determined self-consistently as
\be
I=I_a+I_n=A{e^2 \over 4\pi^2 }\sum_i k_i \mu_i B-{R\over R_0} I={A\over 1+{R\over R_0}}{e^2 \over 4\pi^2 }\sum_i k_i \mu_i  B={I_a\over 1+{R\over R_0}}\quad.\label{totali}
\ee
This is the equation governing the performance of the chiral battery.

Let us now see how the energy discharge works for the chiral battery. From the total current (\ref{totali}) through the resistance $R$ and
the normal current $I_n$ through $R_0$,
the energy discharge rate should be
\be
{d{\cal E}\over dt}=R I^2+R_0 I_n^2={R\over \left(1+{R\over R_0}\right)} A^2 \left({e^2\over 4\pi^2 }\sum_i k_i \mu_i \right)^2 B^2\quad,\label{ratee}
\ee
using (\ref{in}) and (\ref{totali}).
This should match the reduction of internal energy of the Weyl semimetal sample.
In the presence of both electric field $E$ as in (\ref{electric}) and the magnetic field $B$,
the charge density of $i$'th Weyl point changes via triangle anomaly as
\be
{d \rho_i\over dt}={k_i e^2\over 4\pi^2} \vec E\cdot \vec B =-{k_i e^2\over 4\pi^2}{IR\over d}B \quad.
\ee
The total volume of the sample is $Ad$, so that the total rate of increase of $i$'th charge is
\be
{d Q_i\over dt}=-{k_i e^2\over 4\pi^2}AIRB \quad,
\ee
from which the rate of internal energy change is
\be
{d{\cal E}_{\rm int}\over dt} = \sum_i \mu_i {d Q_i\over dt}=-\sum_i {k_i\mu_i e^2\over 4\pi^2}AIRB 
=-{R\over 1+{R\over R_0}}A^2\left({e^2 \over 4\pi^2 }\sum_i k_i \mu_i\right)^2  B^2\quad,
\ee
using the expression (\ref{totali}) for $I$, which indeed agrees precisely with (\ref{ratee}).
The time-dependence of the chiral battery performance relies on the detailed equation of state between $\rho_i$
and $\mu_i$.

 \begin{figure}[t]
 \includegraphics[scale=0.2]{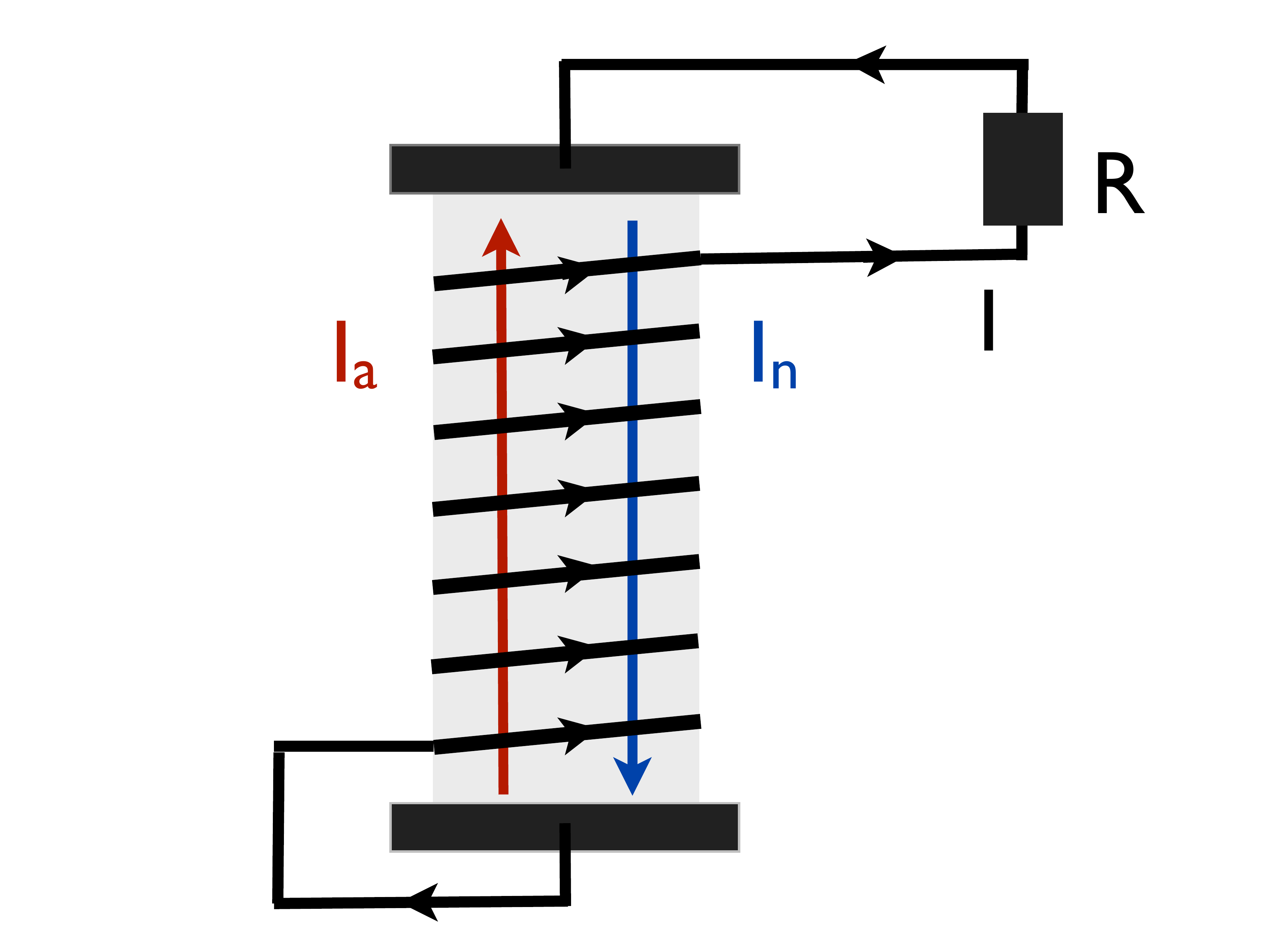}
 \caption{
(color online)  The quantum amplifier: Weyl semimetal (shown in grey) wrapped by a solenoid that is connected to the circuit with resistance $R$. The chiral magnetic anomalous current $I_a$ is generated in response to magnetic field; the resulting voltage drop across the Weyl semimetal sample induces a normal component of the current $I_n$. 
The total current flowing through the resistance is $I = I_a + I_n - dQ/dt$, where $Q$ is the charge accumulated in the capacitor formed by the metal leads attached to the sample (shown in black); note that $I_n$ is negative, see (\ref{In}).
 }
 \label{sol}
 \end{figure}

Although the above example requires an external magnetic field, one can think of other possibility that is completely self-driven, as shown in Fig. 2.
A cylindrical Weyl semimetal sample is put between two metallic plates that form a capacitor with capacitance $C$.
The longitudinal distance and the transverse area of the sample are again $d$ and $A$. In addition
there is a solenoid that wraps the Weyl semimetal sample $N$ times over the distance $d$; this solenoid is connected
to the two metal plates. An external circuit with a resistance $R$ is loaded to the solenoid circuit.
Let the total current along the solenoid and the external resistance $R$ be $I$. 
The strength of the induced magnetic field along $\hat x^3$ inside the solenoid is
\be
B={N\over d} I\quad,\label{B}
\ee
from the Maxwell's equation $\vec \nabla\times \vec B=\vec J$, which gives rise to an anomaly-induced current:
\be
I_a=A J_a= A{e^2\over 4\pi^2 }\sum_i k_i \mu_i  B=\left({e^2\over 4\pi^2 }\sum_i k_i \mu_i\right) {AN\over d} I\quad.\label{Ia}
\ee 
Let the charge stored inside the capacitor formed by the two metallic plates be $Q$. Since the voltage drop along the capacitor
is $\Delta V= Q/C$, there exists a longitudinal electric field (note the negative sign from our definition of $Q$)
\be
E=-{\Delta V\over d}=-{1\over Cd}Q\quad,\label{E}
\ee
and we have a normal current along the Weyl semimetal sample: 
\be
I_n=A\sigma E=-{A\sigma \over C d} Q =-{1\over C R_0} Q\quad,\label{In}
\ee
where $R_0$ is the intrinsic resistance of the sample.
The charge conservation law for the capacitor dictates
\be
{dQ\over dt}=I_a+I_n+I\quad,\label{Qc}
\ee
whereas the Kirchhoff's law of vanishing voltage drop along a closed circuit trajectory gives us
\be
IR+L{dI\over dt} +{1\over C}Q =0\quad,\label{kirch}
\ee
where $L\equiv {A N^2\over d}$ is the inductance of the solenoid.
Using (\ref{Ia}) and (\ref{In}), the (\ref{Qc}) and (\ref{kirch}) take the form
\bear
{dQ\over dt}&=& \left(\left({e^2\over 4\pi^2} \sum_i k_i\mu_i\right){AN\over d} +1\right) I-{1\over R_0 C} Q\quad,\label{dyn1}\\
{dI\over dt}&=& -{R\over L} I-{1\over LC} Q\quad.\label{dyn2}
\eear
These equations comprise a complete dynamical system given
an initial data $(I_0,Q_0)$ at $t=0$, and assuming that the chemical potentials $\mu_i$ are approximately constant during
the discharge process. 
The time evolution of this system can be solved analytically by the Ansatz
\be
\left(\begin{array}{c} Q(t)\\I(t)\end{array}\right)=\left(\begin{array}{c} C_1\\C_2\end{array}\right) e^{\lambda_\pm t}\,,
\ee
where
the characteristic exponents $\lambda_\pm$ of the above system are solutions of
\be
\lambda^2+\left({1\over R_0C}+{R\over L}\right)\lambda+{R\over R_0 LC}+{1\over LC}\left(\left({e^2\over 4\pi^2} \sum_i k_i\mu_i\right){AN\over d} +1\right) =0\quad,
\ee
one of which has a positive real part, and hence {\it instability}, if
\be
-\left({e^2\over 4\pi^2} \sum_i k_i\mu_i\right) > {d\over AN} \left(1+{R\over R_0}\right)\quad.\label{disc}
\ee
When this condition is met, the system has a particular mode in a discharging phase which develops an {\it exponentially increasing current} $I$ (and $Q$).

To see how it can be used for detecting an ultra weak magnetic field, let's introduce a small background external magnetic field $B_{ext}$ to be detected.
The equation (\ref{B}) is modified to be
\be
B={N\over d} I+B_{ext}\quad,\label{BBB}
\ee
so that the above equations of motion (\ref{dyn1}) and (\ref{dyn2}), now including an external magnetic field $B_{ext}$, become
\bear
{dQ\over dt}&=& \left(\left({e^2\over 4\pi^2} \sum_i k_i\mu_i\right){AN\over d} +1\right) I-{1\over R_0 C} Q+A\left({e^2\over 4\pi^2} \sum_i k_i\mu_i\right)B_{ext}\quad,\label{dyn31}\\
{dI\over dt}&=& -{R\over L} I-{1\over LC} Q\quad,\label{dyn3}
\eear
where we have a new source term in the first equation proportional to the external magnetic field. The above inhomogeneous linear differential equation can be solved to give
\be
\left(\begin{array}{c} Q(t)\\I(t)\end{array}\right)=C_+\left(\begin{array}{c} Q_+\\I_+\end{array}\right) e^{\lambda_+ t}+C_-\left(\begin{array}{c} Q_-\\I_-\end{array}\right) e^{\lambda_- t}+
\left(\begin{array}{c} Q_p\\I_p\end{array}\right)\,,
\ee
where $(Q_\pm,I_\pm)^T$ are the eigenvectors of the eigenvalue equation with our previous eigenvalues $\lambda_\pm$,
\be
\left(\begin{array}{cc} -{1\over R_0 C} & \left({e^2\over 4\pi^2} \sum_i k_i\mu_i\right){AN\over d}+1\\-{1\over LC}&-{R\over L}\end{array}\right)\left(\begin{array}{c}Q_\pm\\I_\pm\end{array}\right)=\lambda_\pm \left(\begin{array}{c}Q_\pm\\I_\pm\end{array}\right)\,,
\ee
and $(Q_p,I_p)^T$ is a particular solution sourced by the external magnetic field,
\be
Q_p={\left({e^2\over 4\pi^2} \sum_i k_i\mu_i\right)ARC\over \left({e^2\over 4\pi^2} \sum_i k_i\mu_i\right){AN\over d}+1+{R\over R_0}} B_{ext}\,,\quad I_p=-{1\over RC}Q_p\,.
\ee
The integration constants $C_\pm$ are determined by the initial condition which is naturally $Q(0)=I(0)=0$. For any nonzero $B_{ext}$, one finds that $C_\pm$ are proportional to $B_{ext}$. In particular $C_+\neq 0$ is induced by having $B_{ext}$, so that the mode with $\lambda_+$ which grows exponentially large in time is triggered by the external magnetic field $B_{ext}$.

One can control the instability condition (\ref{disc}) by varying the external resistance $R$ of the circuit. Increasing $R$ beyond a critical value
\be
R>R_c\equiv R_0\left( -\left({e^2\over 4\pi^2} \sum_i k_i\mu_i\right){AN\over d}-1\right)\,,
\ee
will remove the instability from the system, so that the device can stay in a stable condition. One can decrease $R$ below $R_c$ for a detection of an ultra weak magnetic field.

Let us see whether our instability condition (\ref{disc}) can be met with reasonable parameters of the current Weyl semimetals.
We take 5 meV (mili-electron volts) as a typical value for chemical potential of Weyl semimetals, and let's assume a system of a centimeter size.
Note that our formulae are based on the unit system where $\hbar=c=1$, so that the length and energy are related by
$
1\, {\rm cm}^{-1}=2\times 10^{-2}\, {\rm meV}\,,
$
and the fine structure constant is
$
{e^2\over 4\pi}\equiv \alpha={1\over 137}\,.
$
With these values, the left-hand side of (\ref{disc}) is
\be
1.2\times 10^{-2}\left(\sum_i k_i\mu_i\over -5\,{\rm meV}\right) \,{\rm meV}\,,
\ee
whereas the right-hand side of (\ref{disc}) is
\be
1.0\times 10^{-2}\left(d\over 1\,{\rm cm}\right)\left(1\,{\rm cm}^2\over A\right)\left(20\over N\right)\left(1+{R\over R_0}\over 10\right)\,{\rm meV}\,.
\ee 
We see that the condition (\ref{disc}) for the instability can easily be met with a centimeter size of Weyl semimetal device.

\begin{figure}[t]
 \includegraphics[width=5.5cm]{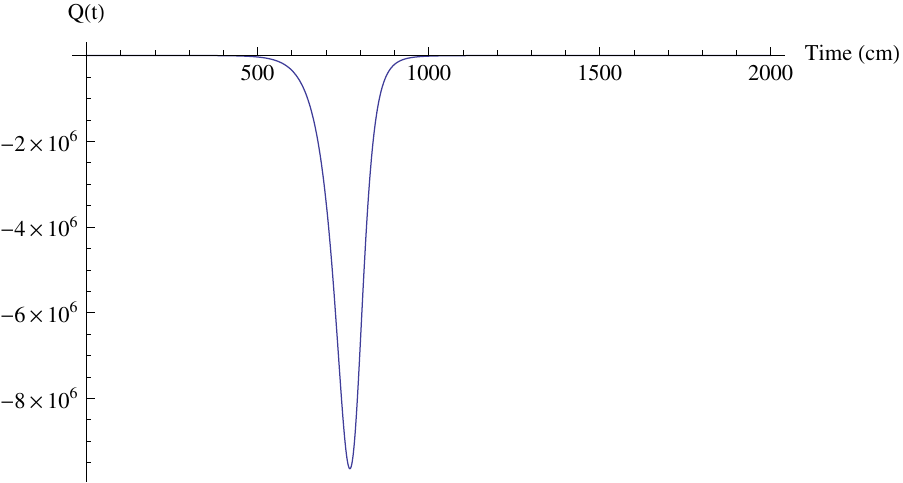} \includegraphics[width=5.5cm]{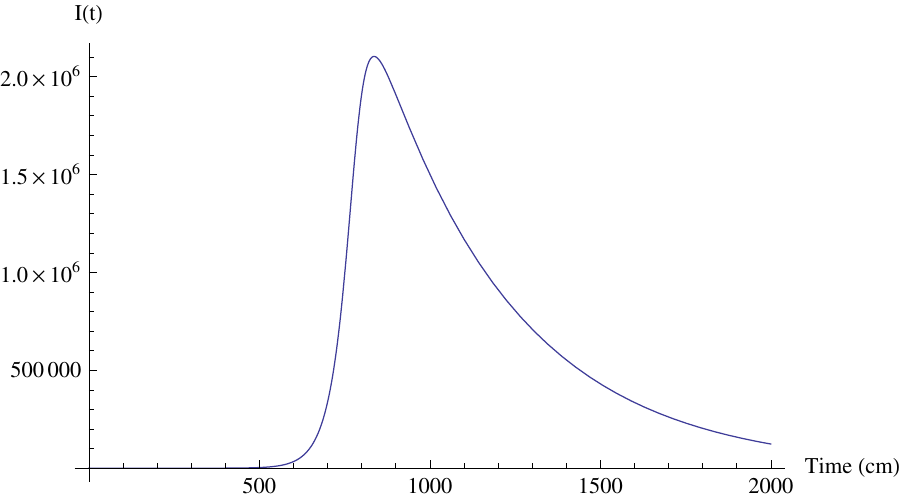} \includegraphics[width=5.5cm]{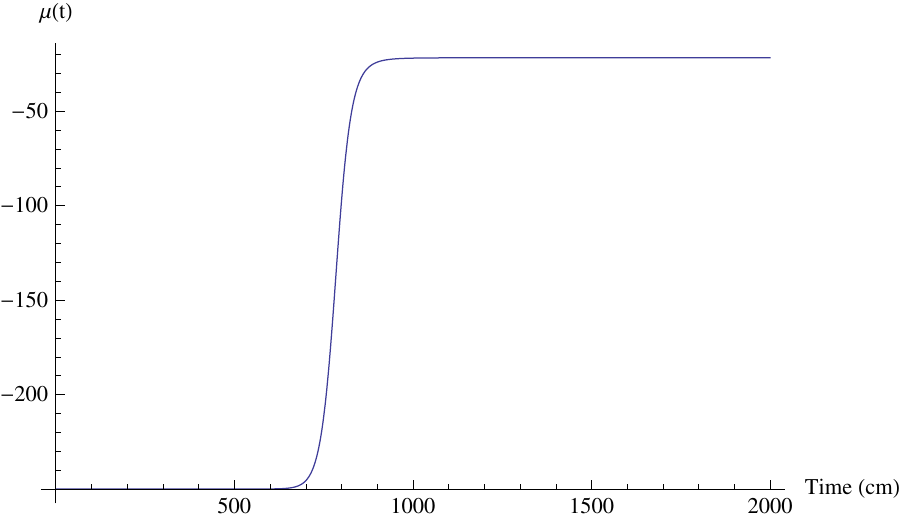}
 \caption{The numerical solution of (\ref{dyn31}), (\ref{dyn3}), and (\ref{dyn4}) with the parameters, $B_{ext}=10^{-8}$ Gauss and $v=0.01$, for a centimeter size device operating at the room temperature. In conventional units, $I=2\times 10^6\,{\rm cm}^{-1}=0.01$ Amperes, and the typical time scale is 1000 cm=30 ns (nanoseconds).
 }
 \label{simulation}
 \end{figure}

As time goes on, one can eventually no longer treat $\mu_i$ constant and the system goes out of the discharging phase (\ref{disc}), and the current $I$ starts to decrease.
This happens because of the triangle anomaly that yields for the $i$'th charge density
\be
{d\rho_i\over dt}={k_i e^2\over 4\pi^2} \vec E\cdot \vec B =-{k_i e^2\over 4\pi^2}\left({N\over C d^2} I Q+{Q\over Cd}B_{ext}\right)\quad,\label{dyn4}
\ee
using (\ref{E}) and (\ref{BBB}) for electric and magnetic fields.
The previous equations (\ref{dyn31}) and (\ref{dyn3}), combined with the above equation (\ref{dyn4}) form a more
complete set of dynamical equations governing the time evolution of the system. 
One needs the equations of state 
\be
\mu_i=\mu_i\left(\{\rho_j\}\right)\quad,
\ee
to solve the system in detail for a concrete realization of the Weyl semimetal.
Around the Weyl points, the dispersion relation of the Weyl excitations is linear in momentum
\be
\epsilon=v|{\bf p}|\,,
\ee
with an effective "speed of light" $v$. In the regime of $\mu\gg T$ (low temperature regime), the density from the Ferm-
Dirac distribution with the above dispersion relation is given by
\be
\rho\approx \int_{|{\bf p}|<p_F}{d^3 p\over (2\pi)^3}={p_F^3\over 6\pi^2}={\mu^3\over 6\pi^2 v^3}\,,\label{eos1}
\ee
using $\mu=vp_F$. In the opposite regime of high temperature $T\gg\mu$, the same computation gives
\be
\rho\approx {T^2\over 6 v^3} \mu\,.\label{eos2}
\ee
The above equations of state (\ref{eos1}) and (\ref{eos2}) can be used to solve the time evolution of the system after our unstable mode is triggered by an external magnetic field.
Since the room temperature $T_r\approx 25$ meV is somewhat larger than $\mu\approx 5$ meV, the latter is suitable for the device application operating in a room temperature.
Let us give an exemplar solution of the time evolution assuming again a centimeter size device operating at the room temperature with a value of $v=0.01$ (in unit of $c$). 
We assume the value of an external magnetic field to be $B_{ext}=10^{-8}$ Gauss=0.5 ${\rm cm}^{-2}$ which is about the magnetic field in the human brain. We also take
the resistance $R_0=R=1$ which are dimensionless in our unit system.
Noting that the capacitance and the inductance are given by 
\be
C={A\over d}=1\,{\rm cm}\,\quad, \quad L={AN^2\over d}=400\,{\rm cm}\,,
\ee
with $A=1\,{\rm cm}^2$, $d=1$ cm, and $N=20$, our equations (\ref{dyn31}), (\ref{dyn3}), and (\ref{dyn4}) become
\be
{dQ\over dt}=(0.046\mu+1)I-Q+1.2\times 10^{-3}\mu\,,\quad {dI\over dt}=-{1\over 400}I-{1\over 400} Q\,,\quad 2.6\times 10^{11}{d\mu\over dt}=-0.046 IQ-1.2\times 10^{-3} Q\,,
\ee
where different quantities are measured in the following units: The electric charge $Q$ is dimensionless, the current $I$ is measured in ${\rm cm}^{-1}$, the chemical potential $\mu$ in ${\rm cm}^{-1}$, time
in cm.
These units can be easily converted into conventional 
ones by restoring the speed of light $c = 3\times 10^{10}$ cm/sec and noting that the charge in Coulombs can be obtained from our (dimensionless) charge by dividing by $6.24 \times 10^{18}$. For example, the typical current on FIG.\ref{simulation} is $I=2\times 10^6\,{\rm cm}^{-1}=0.01$ Amperes, and the typical time scale is 1000 cm=30 ns (nanoseconds).
The initial condition is $I(0)=Q(0)=0$ and $\mu(0)=-5\,{\rm meV}=-2.5\times 10^2\,{\rm cm}^{-1}$.
In FIG. \ref{simulation}, we show the numerical solution of $Q(t), I(t)$ and $\mu(t)$ as a function of time. As we see in the plots, the system initially develops an exponentially large signal
triggered by the external magnetic field before the signal eventually dies out. We have checked numerically that this feature is robust without regard to different values of the parameters.

Even though we need the equation of state to evaluate the detailed time evolution of the system, there are some 
features that are sufficiently general and that in our opinion are quite interesting. Namely, the instability of the circuit of Fig.2 driven by quantum anomaly makes it a {\it quantum amplifier} of magnetic field. In principle, this instability can be induced by a single quantum of magnetic flux through the sample. Therefore 
the considered circuit may be utilized as a sensor of ultra-weak magnetic fields.  
\vskip0.3cm
The circuits discussed above represent only a couple of examples from a vast array of devices that one can envision. We hope that the 
{\it chiral electronics} based on Weyl semimetal circuits can serve as a fascinating way to explore the macroscopic dynamics induced by the 
chiral anomaly, and perhaps open a path towards new electronic devices.


\vskip0.2cm
We thank G. Ba\c sar, G. Dunne, L. Levitov, M. Stephanov, M. \"Unsal and I. Zahed for useful discussions.
This research was supported by the US Department of Energy under Contracts DE-AC02-98CH10886 and DE-FG-88ER41723.



  \end{document}